\documentclass[10pt,a4paper]{article}
\usepackage[utf8]{inputenc}
\usepackage[T1]{fontenc}
\usepackage{amsmath}
\usepackage{amsfonts}
\usepackage{amssymb}
\usepackage{graphicx}
\usepackage{multicol}
\usepackage{float}
\usepackage{dblfloatfix}
\usepackage{placeins}
\usepackage{abstract,lipsum}
\usepackage[left=1.00in, right=1.00in, top=1.00in, bottom=1.00in]{geometry}
\author{Prathyusha S Nair}
\renewenvironment{abstract}
{\quotation\small\noindent\rule{\linewidth}{.5pt}\par\smallskip
	{\bfseries\abstractname\par}\medskip}
{\par\noindent\rule{\linewidth}{.5pt}\endquotation}
\begin{document}
\begin{center}
	\Large{\textbf{Effect of roughness on the transport of an overdamped Brownian particle}}\\[.1in]
	\normalsize {{Prathyusha S Nair$ ^{1}$, Ronald Benjamin$ ^{1} $ }\\[.1in]
		\noindent
		\textit{$ ^{1} $Department of Physics, Cochin University of Science and Technology, Kochi}}\\
	

\end{center}


\begin{abstract}
 We study the transport of an overdamped Brownian particle in a tilted  rough potential driven by a gaussian white noise.We conisder various forms of rough potentials to analyse the effect of roughness on the diffusion and current generated.The variation of current, diffusion coefficient, and peclet number with the static tilting force, amplitude of roughness and temperature were studied. We concluded that the roughness reduces the particle current at any range of temperature. But at low temperatures, it enhances the diffusion coeffeicient of the particle
\end{abstract}
.

\begin{multicols}{2}[\columnsep20pt]

\section{Introduction}
Particle transport by random fluctuations from its environment has implications in various fields of Science. The particles can achieve unidirectional motion by rectifying these random fluctuations. Brownian ratchets are devices that make use of these random fluctuations to move in a particular directions by breaking the detailed balance symmetry. Tilting ratchets, pulsating ratchets, temperature ratchets are some type of ratchets, that has been studied so far. There lies different Physical, chemical and Biological applications of Brownian ratchet mechanism. For example, motion of kinesin motor protein on microtubules, transport of metastatic cancer cells, particle separation and particle pre-concentration based on their size etc.\par
So far, in the study of particle transport smooth type of potentials were considered. But, most of the Brownian particles encounter with a rough environment. When Frauenfelder \textit{et al} studied about the protein folding process \cite{1},\cite{2}, it was understood that this process resulted in some rough/rugged behaviour in the energy landscape of proteins, which affected the movement of proteins. Being inspired from these findings Zwanzig studied about the  diffusion in a rough potential\cite{3}. He generated a rough potential by superimposing a fast randomly oscillating perturbation on a smooth background potential. He proved that the effective diffusion cofficient decreases in the rough potential compared to the smooth potential. Later on, it was also found that the particle current generated decreases in a rough potential, in the overdamped regime\cite{4}.It was  found that when the  noise strength is very low the roughness is found to enhance the particle transport, but when the noise strength is large it hinders the particle transport \cite{5}. The positive effect of roughness on the particle transport was again proved in \cite{6}. On choosing suitbale parameter regime, the roughness can increase the particle current and efficiency.\par
The diffusion of the particle in presence of a rough potential haven't been fully explored.So our aim is to study the behavior of diffusion and directed current in a rough potential.In this paper we are investigating the effect of roughness on the transport of an overdamped  Brownian particle in the presence of a static tilt. We have analysed  transport parameters for different cases of potentials. The model of the system and the potentials we have considered are described in Sec.2. And we discuss about the numerical results in Sec.3. The findings are concluded in Sec.4
\section{Methods}
Here we consider the motion of an overdamped Brownian particle in a periodic rough potential V(x) which is subjected to a static tilting force F and random thermal fluctuations  $\xi$(t). The dynamics of the system is described by the Langevin equation,
\begin{equation}
\gamma \dot{x}= -V'(x) + F + \xi(t)
\end{equation}
The thermal fluctuations are modeled by a gaussian white noise which satisfies,
\begin{equation}
\langle \xi(t)\rangle = 0
\end{equation}
\begin{equation}
\langle \xi(t) \xi(t')\rangle = 2\gamma k\textsubscript{b}T\delta(t-t').
\end{equation}
The periodic potential V(x) contains a smooth part $V_{0}$(x) and a rough part $V_{1}$(x) in the form,
\begin{equation}
V(x) = V_{0}(x) + V_{1}(x)
\end{equation}
We start by considering a sinusoidal function as the smooth potential and roughness also of sinusoidal form(Case I).
\begin{equation}
V_{0}(x)= sin(2 \pi x/L)
\end{equation} 
\begin{equation}
V_{1}(x)= \epsilon (c_{1}\sin (\lambda_{1}\Omega x) +c_{2}\sin (\lambda_{2}\Omega x)).
\end{equation}
Here L is the spatial period of the potential $V_{0}(x)$ and $\epsilon$ is the amplitude of roughness. $\lambda_{1}\Omega$, $\lambda_{2}\Omega$ and  $c_{1}$,$c_{2}$ are the periodicities and amplitudes of the two sine functions in $V_{1}$. Higher the values of $\lambda$s, denser the perturbations. The perturbation becomes deeper on increasing the $\epsilon$. The values of parameters ,$c_{1}$,$c_{2}$,$\lambda_{1}$, $\lambda_{2}$ are taken as 1, 0.5, 1, and 2 respectively. The spatial period L is taken as 2$\pi$.\par
Next we consider the case, where the unperturbed  potential is a spatially asymmetric function.(Case II). The potential $V_{0}(x)$ takes the form,
\begin{equation}
V_{0}(x)= - [\sin(x) + \frac{1}{4}\sin(2x + \psi - \frac{\pi}{2})] 
\end{equation}
The rough part of the  potential is given by, equation (6). Here $\psi$ is the parameter that breaks the symmetry of the potential.The potential is periodic with period L =2$\pi$.\par
As a next case (Case III ) a piece wise linear potential is taken for analysis.
\begin{equation}
V_{0} =	\begin{cases}
\frac{x}{\alpha} ,&0\le x < \alpha\\
\frac{\lambda-x}{\lambda-\alpha},&\alpha <x<\lambda
\end{cases}
\end{equation}
where $\alpha$ is the asymmetry parameter and $\lambda$ is the spatial period of the system. When $\alpha$=0.5 then, the potemtial is symmetric in space. Here a regular sinusoidal potential is considered as the roughness.
Therefore the total potential will be,
\begin{equation}
V(x)=[V_{0}(x) - \epsilon\cos(2\pi kx)/2]/N
\end{equation}
where k  is the wavenumber which is an odd integer and N is the normalization factor. Here we have taken N as (1+$\epsilon$).\par
The effective transport of a Brownian particle is defined by the coherency of the system. The more coherent the particles are the more effecyive is the transport. So here we study some parameters that measures the effectiveness of the transport.
The quantity of primary interest is the particle current generated. It is defined as,

\begin{equation}
\langle \dot{x}\rangle = \lim_{t \to \infty} \frac{\langle x(t)\rangle}{t}.
\end{equation}
The  brackets $\langle$..$\rangle$ denotes the ensemble average of the quantity. We then study how much the particle trajectory has spread. This is quantified by the effective diffusion coefficient, which is given by,

	\begin{equation}
D\textsubscript{eff} = \lim_{t \to \infty}\frac{\langle x^2(t)\rangle -\langle x(t) \rangle ^2}{2t}.
\end{equation}
 The Peclet number is a transport parameter that incoperates velocity of the ratchet and coherency of the ratchet. It is described as,
 \begin{equation}
 Pe = \frac{L \langle v \rangle}{D_{eff}}
 \end{equation} where v is the velocity of the particle, $D_{eff}$ is the effective diffusion coefficient. L is the characteristic length of the sysytem. Generally we take the spatial period of the  potential as the characteristic length.\par
  The effective diffusion coefficient for a particle in a periodic potential . \cite{7},\cite{8},\cite{9} is defined according to the Refs as,
  \begin{equation}
  D_{eff} = \frac{D_{O} \int_{x_{0}}^{x_{0}+L} I_{+}(x)[I_{-}(x)]^2 \frac{dx}{L}}{[\int_{x_{0}}^{x_{0}+L} I_{-}(x)\frac{dx}{L}]^3} 
  \end{equation}
  where,
  \begin{equation}
  I_{\pm}(x)=\mp D_{0}^{-1} e^{ \pm V(x)/k_{b}T}\int_{x}^{x\mp L}e^{\mp V(x)/k_{b}T} dy
  \end{equation}
  in which $x_{0}$ is an arbitrary position and $D_{0}= k_{b}T/\gamma$. And the particle current of the system is equal to \cite{7,8,9},
  \begin{equation}
 \langle \dot{x}\rangle = N^{-1}(1 - e^{-LF/k_{b}T})
  \end{equation}\par
 Here we have carried out numerical simulations to calculate the transport properties. Equation (1) has been numerically integrated using Euler's method with time step dt $\le$ $10^{-4}$ for Case I and Case III. The Case II was simulated using stochastic Runge kutta method of order two with time step dt $\le$ $10^{-4}$ .  The total time of observation was taken as 1000 units. The initial position of the particle was randomly chosen from a uniform distribution in the interval [0,1]. The transport properties were calculated by averaging over 1000 trajectories.
 
%
%
%
 
%
 
 
\section{Results and Discussions}
In this study we have considered different types of potentials to analyse the effect of roughness on the system. Lets take the Case I. To investigate how the roughness changes the transport proprties, we first analyse the particle current as a function of tilting force for different values of amplitude of roughness $\epsilon$. In Fig 1a and Fig.2 we can see that the current increases monotonically as a function of F. Increasing the tilting force F, helps the particle to over come the potential barrier. To analyse the effect of roughness we focus on to the values of F upto 10(Fig. 1.b, Fig.3). We can see that roughness doesnot contribute to the transport. The particle current is high when there is no roughness. The current decreases with the roughness $\epsilon$. Fig 4.also affirms this observation. For different temperatures,increasing  the roughness seems to hinder the particle current. When the amplitude of roughness increases it becomes difficult for the particle to overcome each potential minima due to the rough part. 
 \begin{figure*}[t]
	\includegraphics{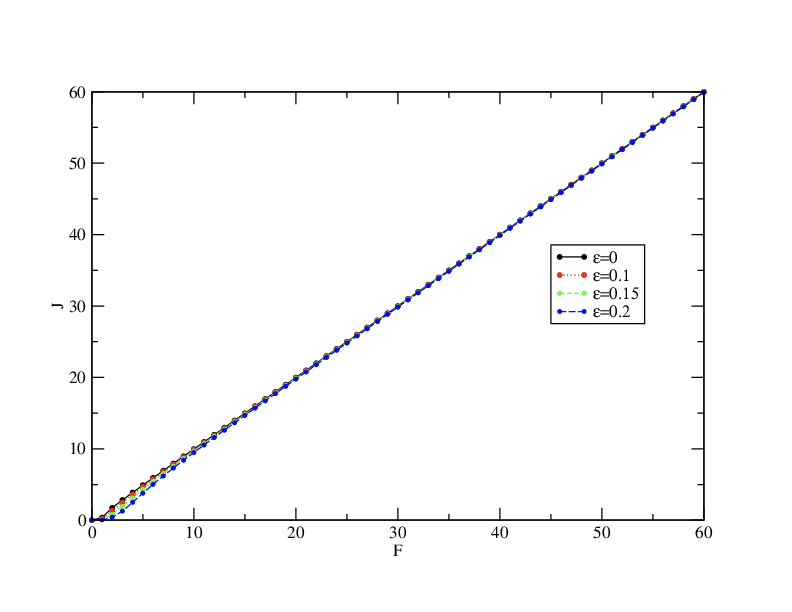}
	\includegraphics{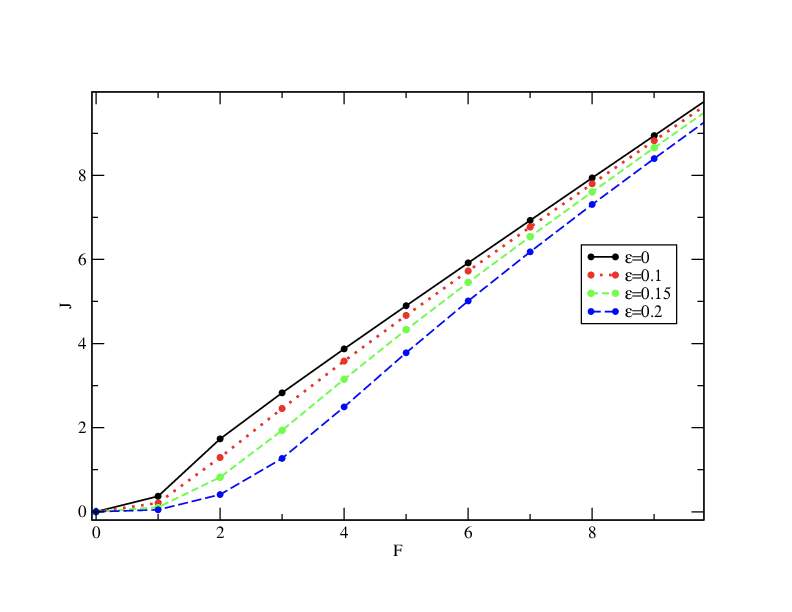}
	\caption{a) The particle current J as a function of Tilting force for various values of $\epsilon$ at temperature T=0.1 and $\lambda_{1}$=10 $\lambda_{2}$=20. b)J as a function of F  where F$\le$ 10(Case -I)}
\end{figure*}
%

 And when we vary the temperature, the current is found to increases with it(see Fig. 5). In the absence of roughness the J is approximately same for all values of Temperature. When the temperature is very much lower than the value of $\epsilon$ (less than 0.05), the current J is negligible in presence of roughness. With such a low value of temperature it seems difficult for the particle cross the potential minima. The temperature determines the noise strength. So only in presence of higher noise strength the particle get enough pull to easily move out of the potential well.\par

 \begin{figure}[H]
 	\includegraphics[scale=0.3]{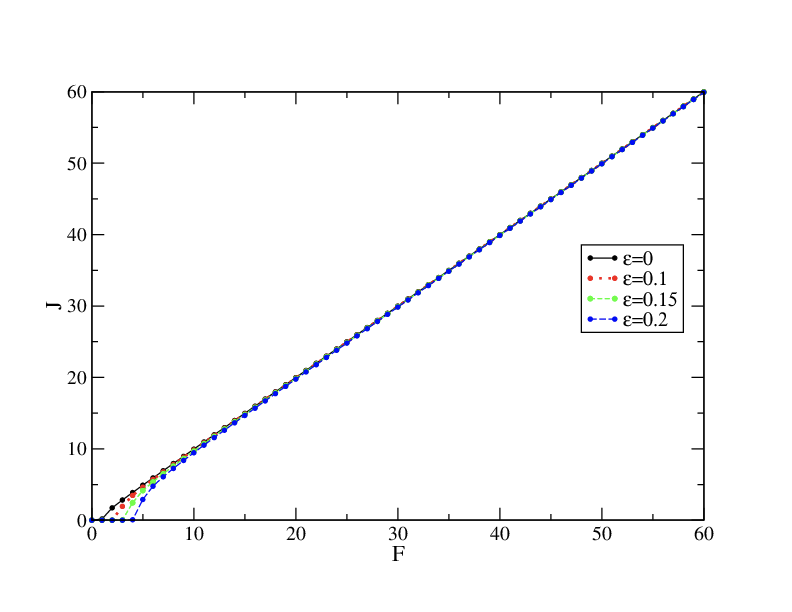}
 	\caption{The particle current J as a function of Tilting force F for various values of $\epsilon$ at temperature T=0.01 and $\lambda_{1}$=10 $\lambda_{2}$=20(Case -I)}
 \end{figure}
 \begin{figure}[H]
 	\includegraphics[scale=0.3]{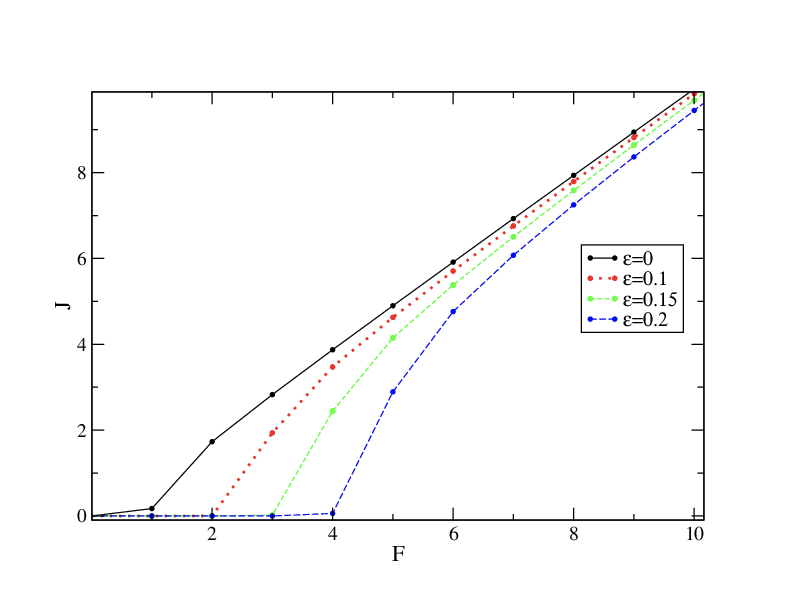}
 	\caption{The particle current J as a function of Tilting force F for various values of $\epsilon$ at temperature T=0.01 and $\lambda_{1}$=10 $\lambda_{2}$=20 for smaller values of F.(Case -I)}
 \end{figure}
The effective diffusion coefficient is numerically calculated using equation(11). The effective diffusion coefficient is high when the particle diffuses over the potential barrier. In Fig.6 and Fig.7 we have studied the variation of diffusion coefficient as a function of the tilting force F at two different temperatures.

\begin{figure}[H]
	\includegraphics[scale=0.3]{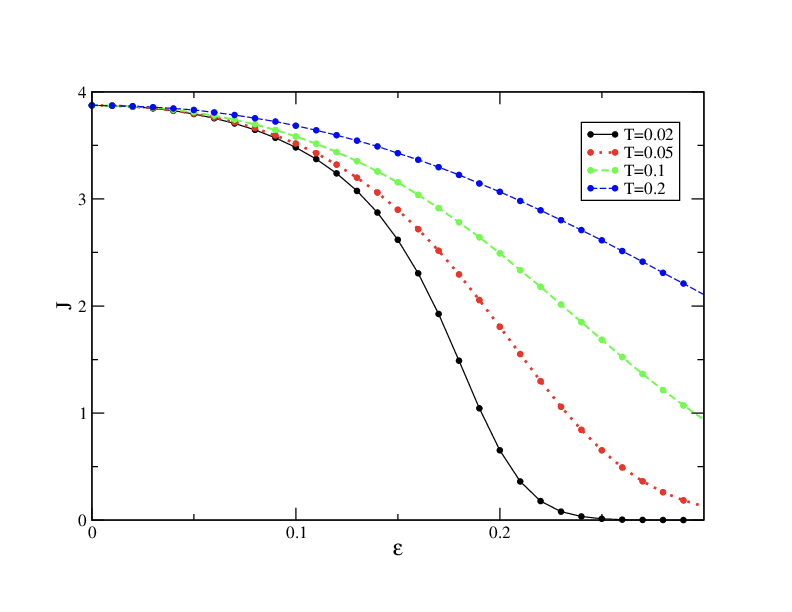}
	\caption{ The particle current as a function of the amplitude of roughness $\epsilon$ for various Temperatures when the tilting force F =4 and  $\lambda_{1}$=10 $\lambda_{2}$=20.(Case-I)}
\end{figure}
\begin{figure}[H]
	\includegraphics[scale=0.3]{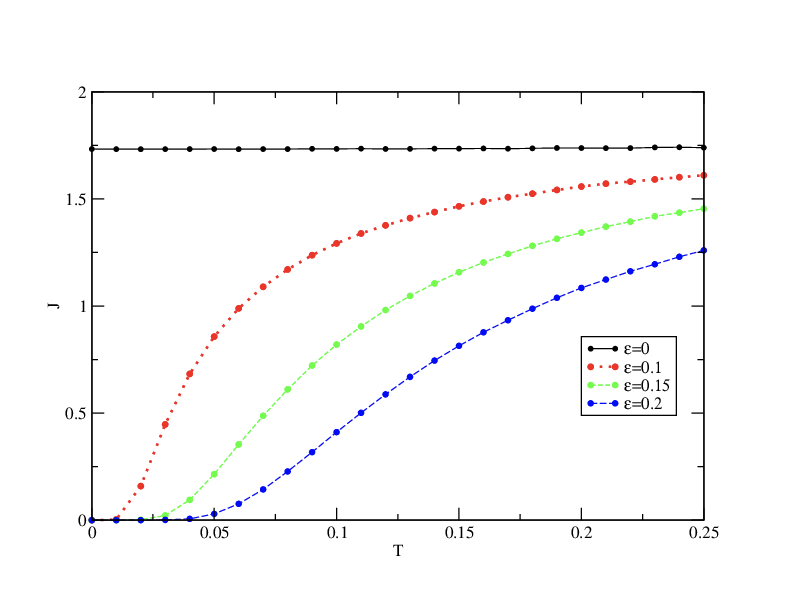}
	\caption{The particle current J as a function of temperature T for various values of $\epsilon$ when tilting force F=2 and $\lambda_{1}$=10 $\lambda_{2}$=20. (Case-I)}
\end{figure}

 	\begin{figure}[H]
 		\includegraphics[scale=0.33]{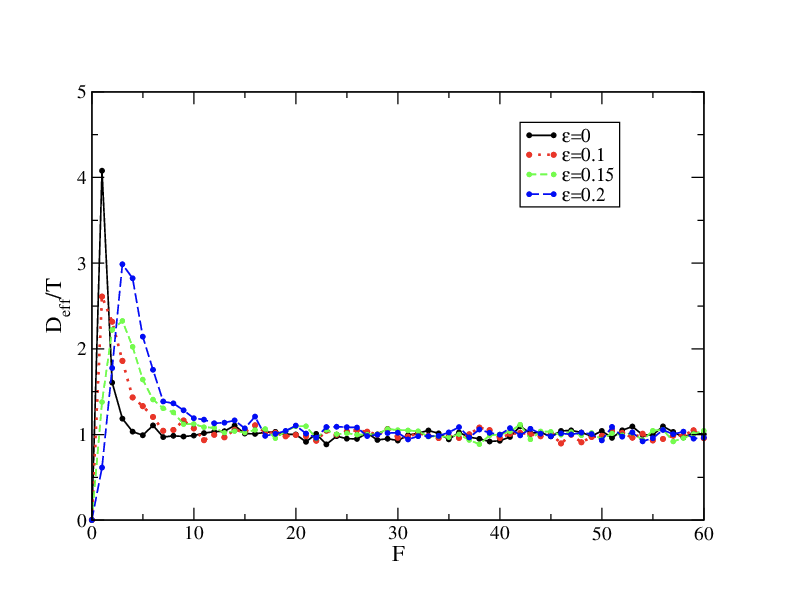}
 		\caption{The Effective diffusion coefficient as a function of tilting force F for different values of $\epsilon$ at a temperature T =0.1 and $\lambda_{1}$=10, $\lambda_{2}$=20. (Case-I)}
 	\end{figure}
 	\begin{figure}[H]
 		\includegraphics[scale=0.33]{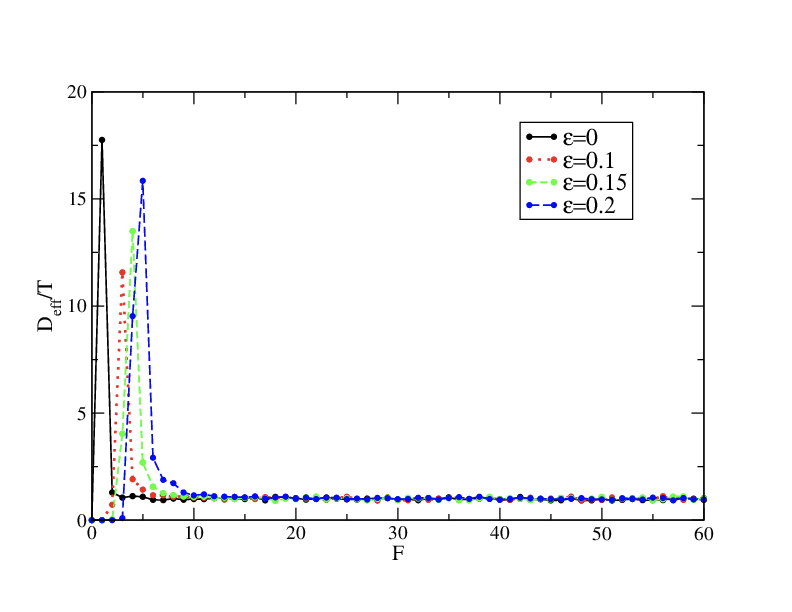}
 	\caption{The Effective diffusion coefficient as a function of tilting force F for different values of $\epsilon$ at a temperature T =0.01 and $\lambda_{1}$=10, $\lambda_{2}$=20. (Case-I)}
 	\end{figure}

	\begin{figure}[H]
		\includegraphics[scale=0.3]{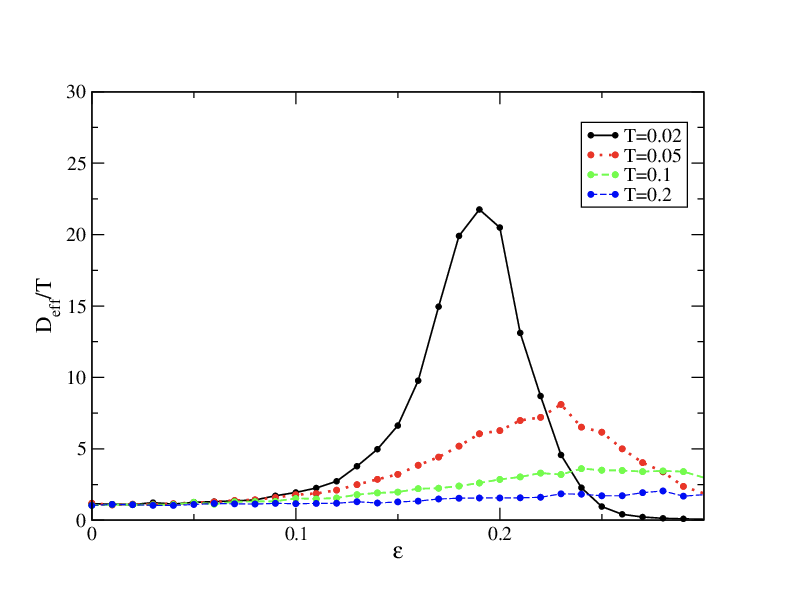}
		\caption{The Effective diffusion coefficient as a function of $\epsilon$ for different values of temperature when the tilting force F =4 and $\lambda_{1}$=10, $\lambda_{2}$=20.(Case-I)}
	\end{figure}
	\begin{figure}[H]
		\includegraphics[scale=0.3]{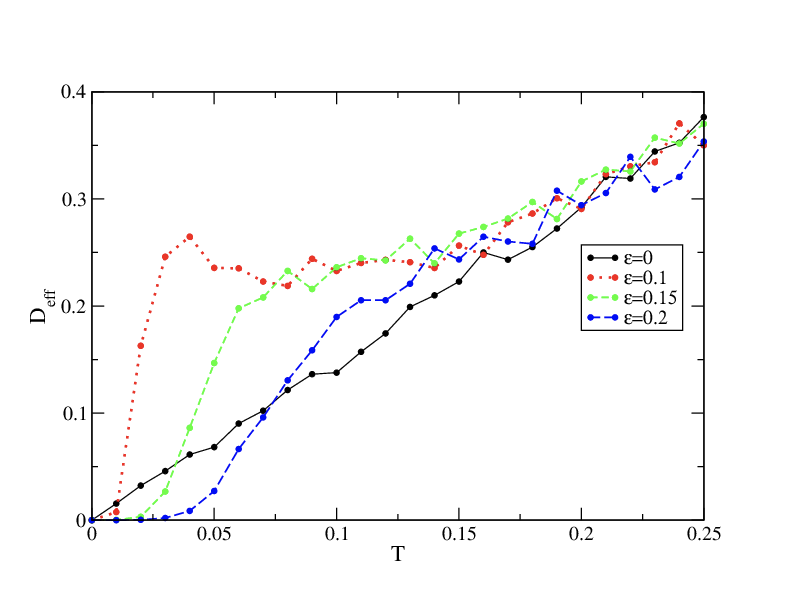}
		\caption{The Effective diffusion coefficient as a function of temperature T for various values of $\epsilon$ when tilting force F=2 and $\lambda_{1}$=10 $\lambda_{2}$=20.(Case-I)}
	\end{figure}
When the tilting force is greater than zero (F>0), slope of the effective potential is negative and the particle moves in the positive direction. At a critical value of force F it over comes the force due to the potential. And the particle diffuses out from the potential minima. The peaks in the curve suggests that the tilting force F has over come the force due to the potential barrier. We can see that till a particular value of F the roughness seems to hinder the particle diffusion. And after that point the roughness is enhancing the effective diffusion coefficient upto a value of F. The roughness does not seem to have much effect on $D_{eff}$. When the temperature decreases the position of critical values of F shifts to the left. The observation we had about the effect of rouhness on  $D_{eff}$is affirmed in Fig.8. 
	\begin{figure}[H]
		\includegraphics[scale=0.33]{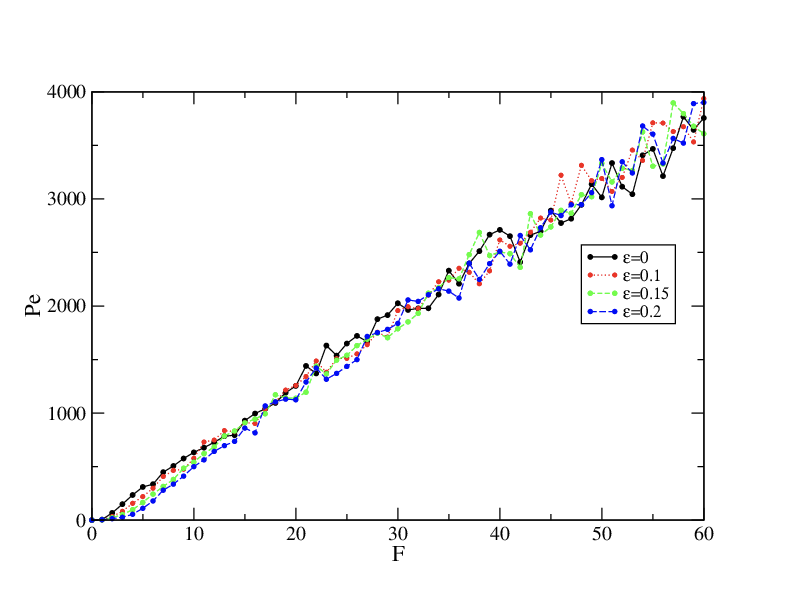}
		\caption{The Peclet number as a function of tilting force F for different values of $\epsilon$ at a temperature T =0.1 and $\lambda_{1}$=10, $\lambda_{2}$=20. (Case-I)}
	\end{figure}
	\begin{figure}[H]
		\includegraphics[scale=0.33]{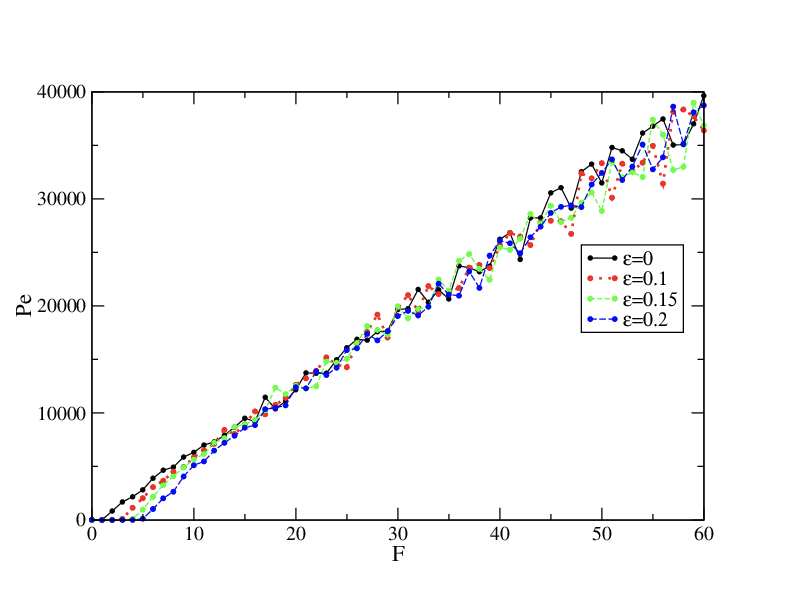}
		\caption{The Peclet number as a function of tilting force F for different values of $\epsilon$ at a temperature T =0.01 and $\lambda_{1}$=10, $\lambda_{2}$=20. (Case-I)}
	\end{figure}

	\begin{figure}[H]
		\includegraphics[scale=0.33]{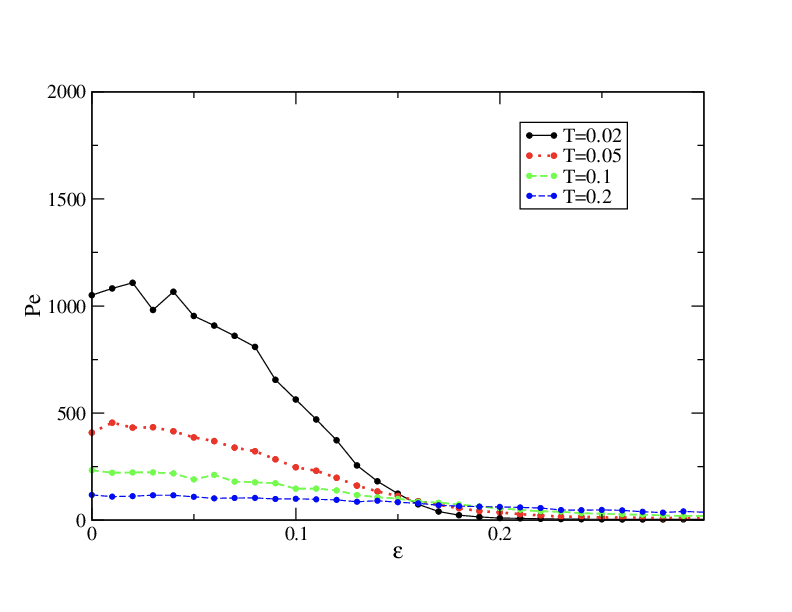}
		\caption{The Peclet number Pe as a function of $\epsilon$ for different values of temperature when the tilting force F =4 and $\lambda_{1}$=10, $\lambda_{2}$=20.(Case-I)}
	\end{figure}
	\begin{figure}[H]
		\includegraphics[scale=0.33]{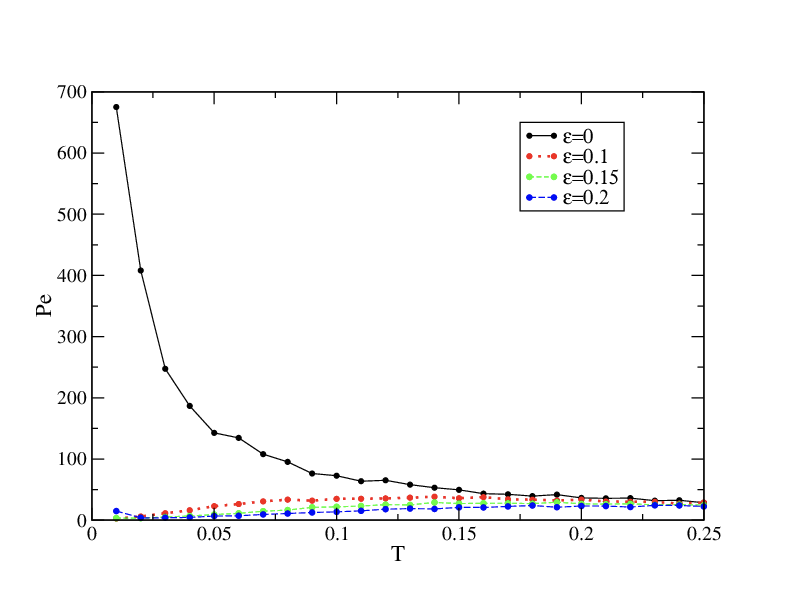}
		\caption{The Peclet number Pe as a function of temperature T for various values of $\epsilon$ when tilting force F=2 and $\lambda_{1}$=10 $\lambda_{2}$=20.(Case-I)}
	\end{figure}

At very low temperatures the effective diffusion coefficient increases upto a particular value of roughness and then it decreases. And the roughness seems to have negligible effect on Diffusion coefficient at higher temperatures. This is again clarified by varying Diffusion coefficient as a function of temperature Fig.9. As we increase the temperature we find that the diffusion coefficient is almost the same for different values of roughness.

Now we move to \textbf{Case II}. Here we have taken the asymmetric case by choosing the aymmetry parameter as $\psi = 0.5\pi$. The parameters of rough part (Equation 7) $c_{1}$,$c_{2}$, $\lambda_{1}$, $\lambda_{2}$ are chosen as 1, 0.5, 1,and 2. We have analysed particle current as a function of tilting force for different values of $\epsilon$ at different temperatures and different $\Omega$s (Fig.14, Fig.15, Fig.16). As we increase the tilting force the effective potential becomes more and more tilted, which helps the particle to overcome the potential barrier easily.Here the slope of the potential is negative and hence the particle moves in the positive direction. It is evident fro thee figures that, the roughness reduces the particle current for smaller vaues fo tilting force. But at higher values of F, roughness doesnot seem to effect the particle cuurent. The particle current is almost same for various values of roughness.

%

%
%
%

 		\begin{figure}[H]

	\includegraphics[scale=0.3]{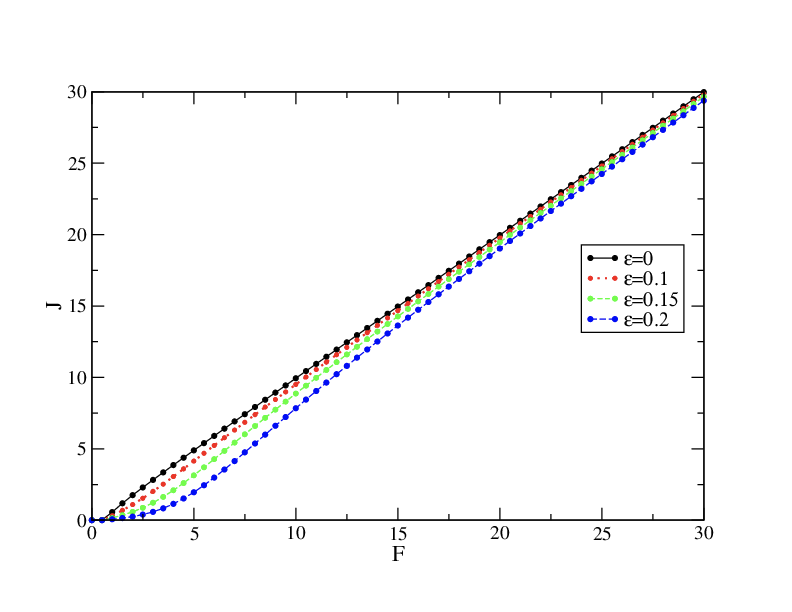}
	\caption{ The particle current J as a function of Tilting force for various values of $\epsilon$ at temperature T=0.1 and  $\Omega=20$ .(Case -II)}
\end{figure}
%
\begin{figure}[H]
	
	\includegraphics[scale=0.3]{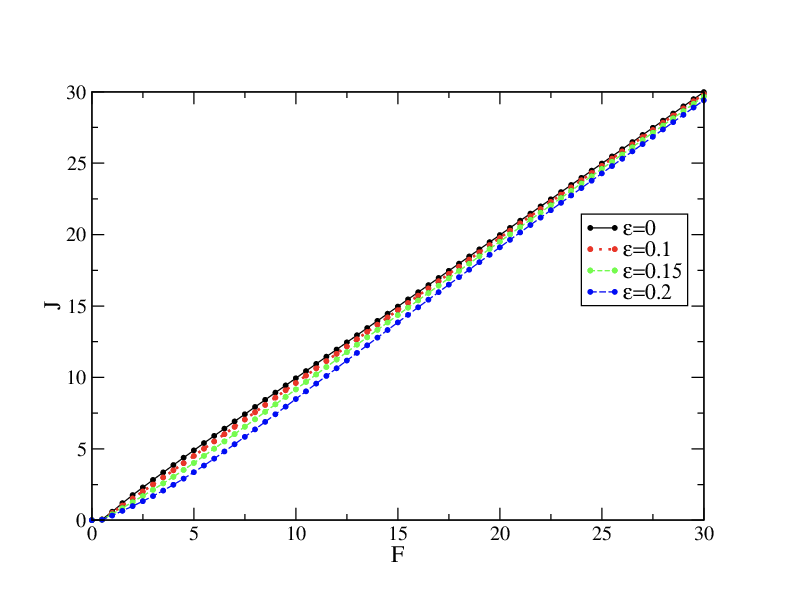}
	\caption{ The particle current J as a function of Tilting force for various values of $\epsilon$ at temperature T=0.2 and  $\Omega=20$ .(Case -II)}
\end{figure}

The effect of roughness on the particle current is visible from Fig.18 and Fig.19. As we increase the amplitude of roughness, it becomes difficult for the particle to jump out of these  minima of the rough potential. At higher temperatures the particle current will be higher. And when we increase the tilting force, the maximum value of the current increases.\par
 To study the behaviour of diffusion in the second case we varied the diifusion coefficient as a function of the static tilting force for different amplitude of roughness at different temperatures(Fig.21,22,23). The diffuion coefficient $D_{eff}$ increases with the tilting force upto a critical value and then it decreases. In the rough potential the tilting force has to overcome the force due to the smooth part and force due to the rough part of the potential. Hence there  occurs multiple peaks in the graph each time the particle moves out of the potential minima.

%

\begin{figure}[H]
	
	\includegraphics[scale=0.3]{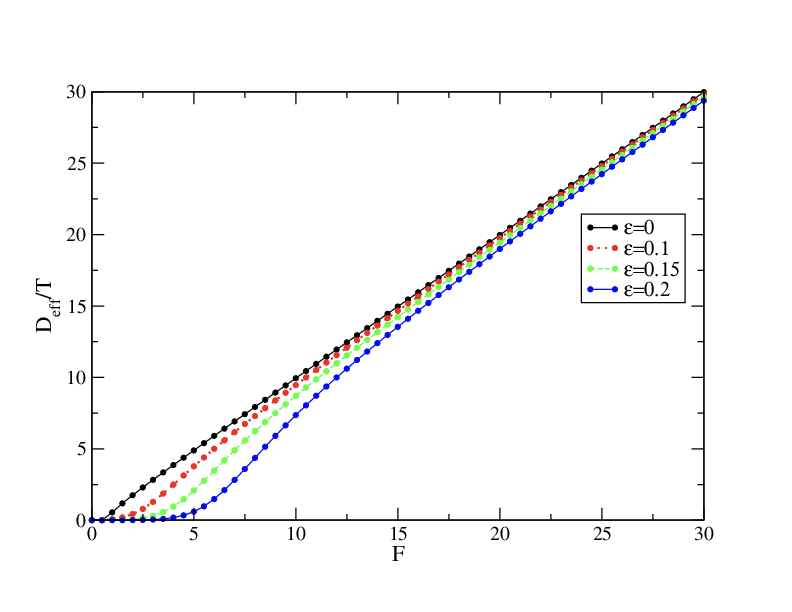}
	\caption{ The particle current J as a function of Tilting force for various values of $\epsilon$ at temperature T=0.05 and $\Omega=20$ .(Case -II)}
\end{figure}

%
\begin{figure}[H]

\includegraphics[scale=0.33]{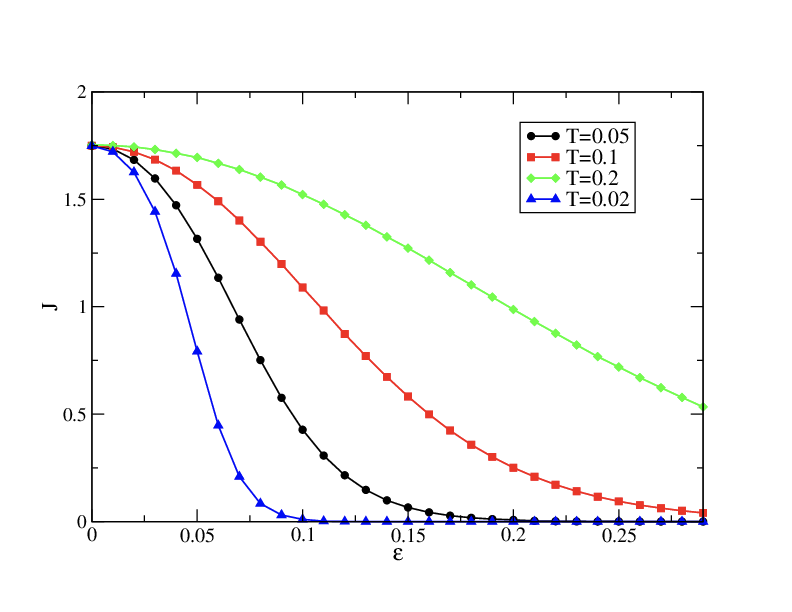}
\caption{The particle current J as a function of $\epsilon$ for different values of temperature when the tilting force F =2 and  $\Omega$ =20 (Case-II).}
\end{figure}
%
%
\begin{figure}[H]
	
	\includegraphics[scale=0.33]{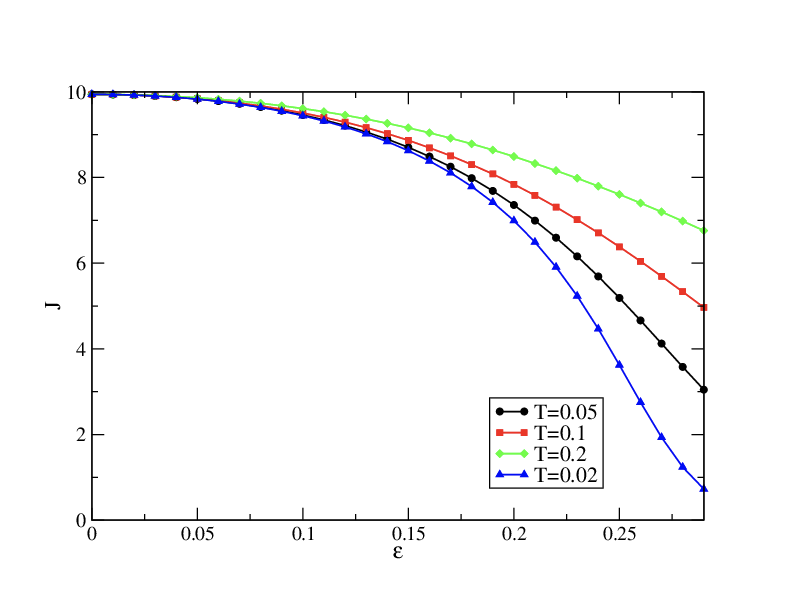}
	\caption{The particle current J as a function of $\epsilon$ for different values of temperature when the tilting force F =10 and $\Omega$ =20 (Case-II).}
\end{figure}
\begin{figure}[H]
	\includegraphics[scale=0.33]{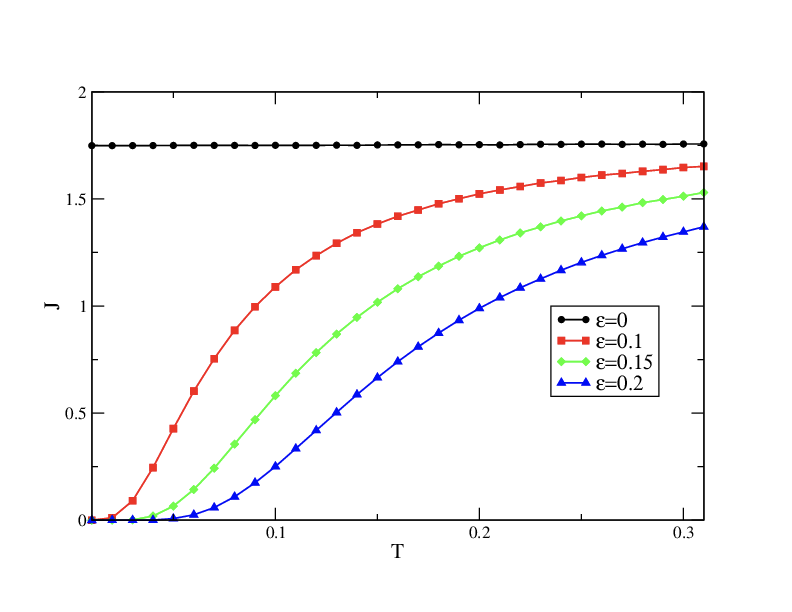}
	\caption{The particl current J as a function oftemperature T   for different values of  $\epsilon$ at F=2 with  $\Omega=20$}
\end{figure}
\begin{figure}[H]
	\includegraphics[scale=0.33]{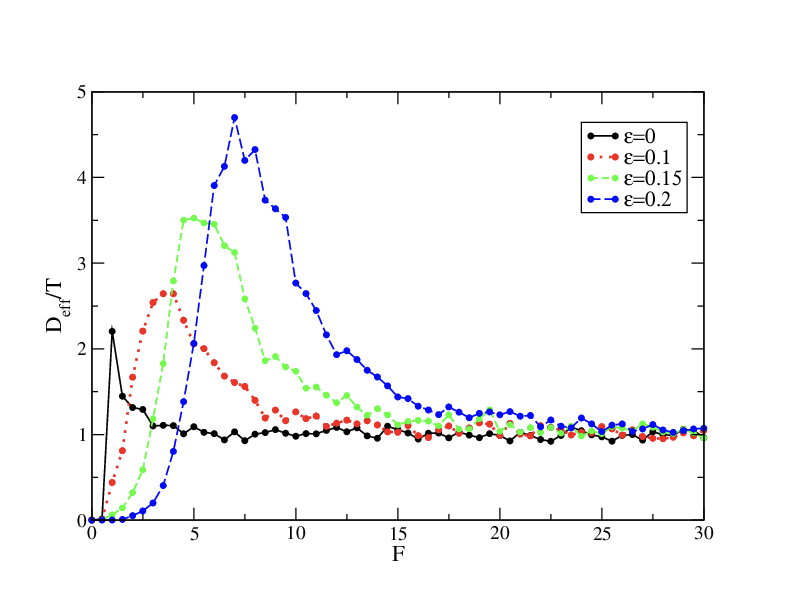}
	\caption{The effective diffusion coefficient as a function of tilting force F for different values of $\epsilon$ at temperature T=0.05 with  $\Omega=20$}
\end{figure}
\begin{figure}[H]
\includegraphics[scale=0.33]{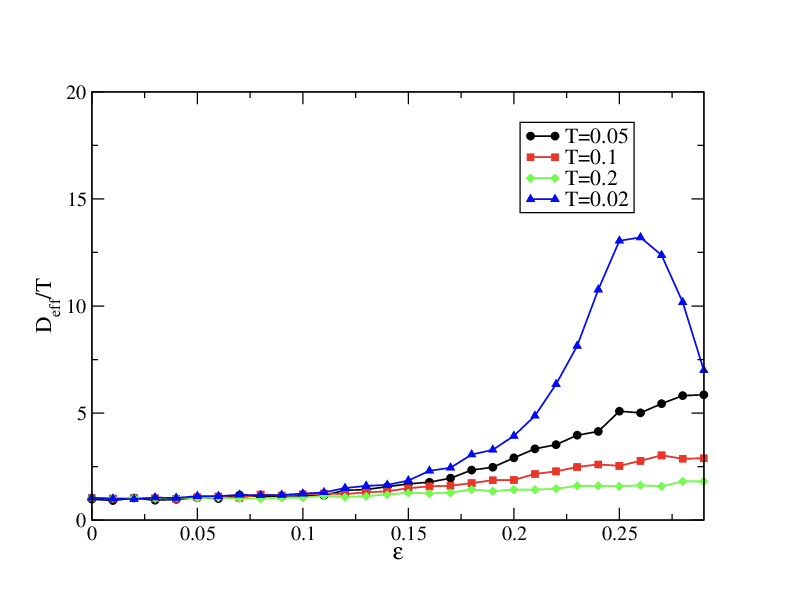}
\caption{The effective diffusion coefficient as a function of amplitude of roughness $\epsilon$  for different values of  temperature T at F=10 with  $\Omega=20$}
\end{figure}
\begin{figure}[H]
	\includegraphics[scale=0.33]{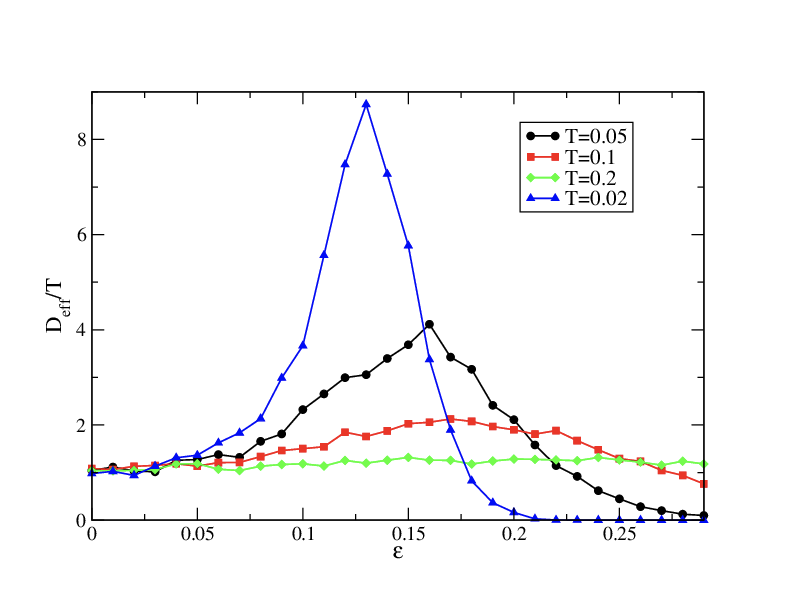}
	\caption{The effective diffusion coefficient as a function of amplitude of roughness $\epsilon$  for different values of  temperature T at F=5 with  $\Omega=20$}
\end{figure}
\begin{figure}[H]
	\includegraphics[scale=0.33]{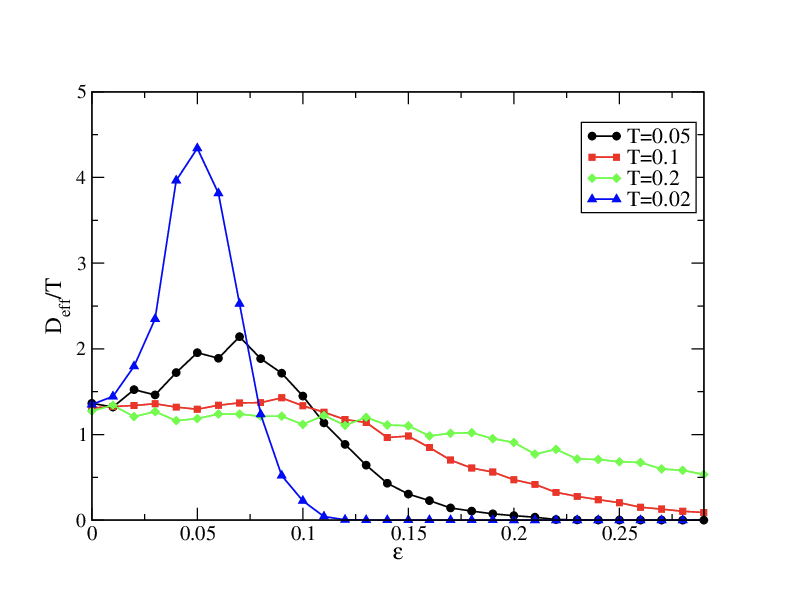}
	\caption{The effective diffusion coefficient as a function of amplitude of roughness $\epsilon$  for different values of  temperature T at F=2 with  $\Omega=20$}
\end{figure}
\begin{figure}[H]
	\includegraphics[scale=0.33]{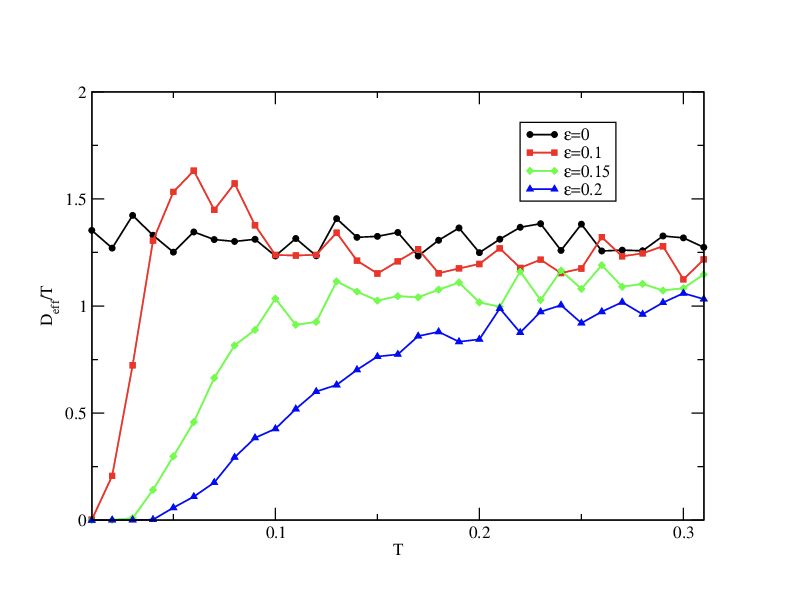}
	\caption{The effective diffusion coefficient as a function of temperature T   for different values of  $\epsilon$ at F=2 with  $\Omega=20$}
\end{figure}
%

\begin{figure}[H]
	\includegraphics[scale=0.33]{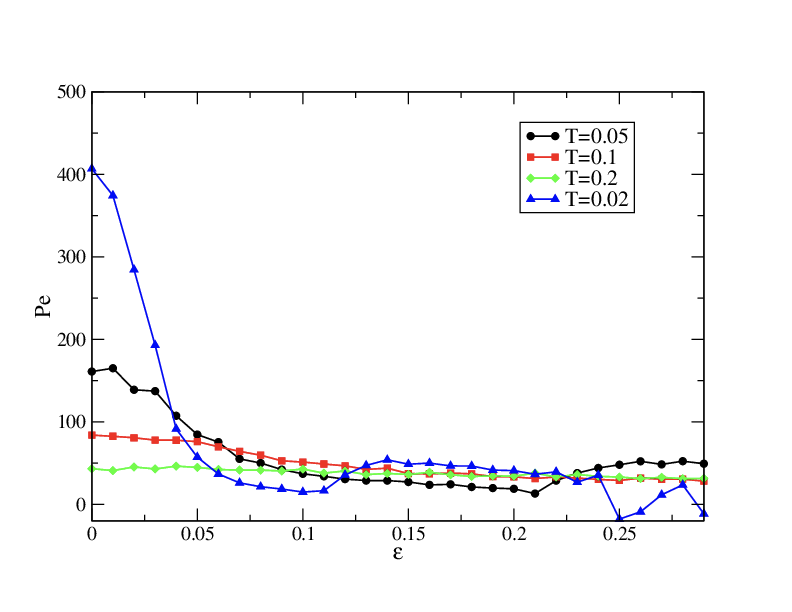}
	\caption{The Peclet number as a function of amplitude of roughness $\epsilon$  for different values of  temperature T at F=2 with  $\Omega=20$}
\end{figure}
\begin{figure}[H]
	\includegraphics[scale=0.33]{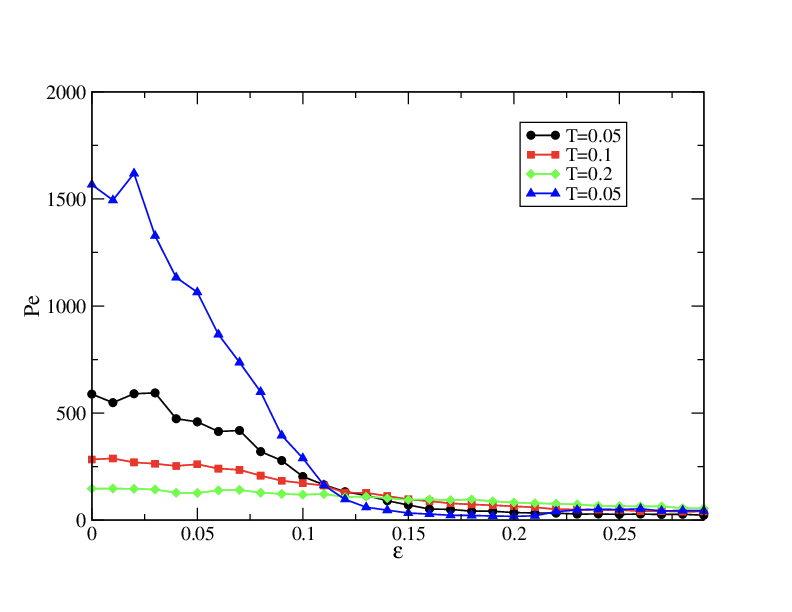}
	\caption{The Peclet number as a function of amplitude of roughness $\epsilon$  for different values of  temperature T at F=5 with  $\Omega=20$}
	\end{figure}
\begin{figure}[H]
	\includegraphics[scale=0.33]{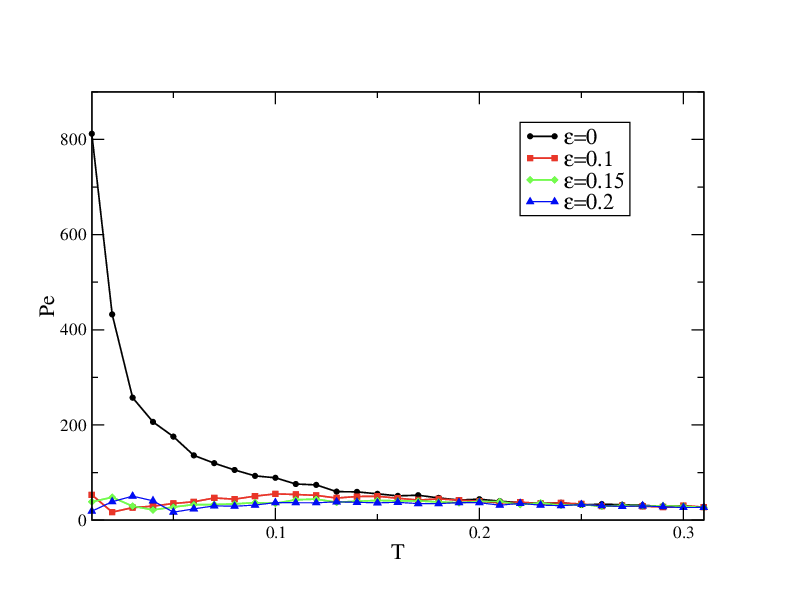}
	\caption{The Peclet number as a function of temperature T   for different values of  $\epsilon$ at F=2 with  $\Omega=20$}
	\end{figure}
\section{Conclusions}
On analysing the behaviour of current in various potential, we found that, the roughness does have a negative impact on the particle current. The particle current was found to decrease with roughness  at all range of temperatures we have considered. But at low temperatures the diffusion of the particle is being enhanced by the roughness. The diffusion coefficient of the particle is found to increase with roughness after a critical value of F, at lower temperatures. At higher temperatures roughness doesn't effect the effective diffusion coefficient. So we concluded that roughness does not promote directed transport at lower temperatures.
 
\end{multicols}
\end{document}